\documentclass[authoryear,preprint,12pt]{elsarticle}
\usepackage{subfigure}
\usepackage{amssymb}
\usepackage{amstext}
\usepackage{lscape}
\journal{Radiation Physics and Chemistry}
\begin{document}
\begin{frontmatter}
  \title{Reconstruction of bremsstrahlung spectra from attenuation data using generalized simulated annealing}
  \author[ffclrp,ifsp]{O. H. Menin}
	\author[ffclrp,inctsc]{A. S. Martinez}
	\author[ffclrp,inctmrm]{A. M. Costa\corref{cor1}}
  \ead{amcosta@usp.br}
  \address[ffclrp]{Departamento de F\'{i}sica, Faculdade de Filosofia, Ci\^{e}ncias e Letras de Ribeir\~{a}o Preto, Universidade de S\~{a}o Paulo, Av. Bandeirantes 3900, 14040-901, Ribeir\~{a}o Preto, SP, Brazil}
	\address[ifsp]{Instituto Federal de Educa\c{c}\~{a}o, Ci\^{e}ncia e Tecnologia de S\~{a}o Paulo, Rua Am\'{e}rico Ambr\'{o}sio 269, 14169-263, Sert\~{a}ozinho, SP, Brazil}
	\address[inctsc]{Instituto Nacional de Ci\^encia e Tecnologia em Sistemas Complexos, Brazil}
  \address[inctmrm]{Instituto Nacional de Ci\^encia e Tecnologia em Metrologia das Radia\c{c}\~oes em Medicina, Brazil.}
  \cortext[cor1]{Corresponding author. Telephone: +55-16-3602-4670 Fax: +55-16-3602-4887.}
  \begin{abstract}
    The throughout knowledge of a X-ray beam spectrum is mandatory to assess the quality of its source device. Since the techniques to directly measurement such spectra are expensive and laborious, the X-ray spectrum reconstruction using attenuation data has been a promising alternative. However, such reconstruction corresponds mathematically to an inverse, nonlinear and ill-posed problem. Therefore, to solve it the use of powerful optimization algorithms and good regularization functions is required. Here, we present a generalized simulated annealing algorithm combined with a suitable smoothing regularization function to solve the X-ray spectrum reconstruction inverse problem. We also propose an approach to set the initial acceptance and visitation temperatures and a standardization of the objective function terms to automatize the algorithm to address with different spectra range. Numerical tests considering three different reference spectra with its attenuation curve are presented. Results show that the algorithm provides good accuracy to retrieve the reference spectra shapes corroborating the central importance of our regularization function and the performance improvement of the generalized simulated annealing compared to its classical version.
  \end{abstract}
\end{frontmatter}

\section{Introduction}
\label{introduction}

A diagnostic  X-ray beam can be characterized by its photon fluence distribution, which corresponds to the number of photons within each energy interval in the  spectrum range. Experimentally, this distribution is generally determined by direct measurement through the use of an energy dispersive solid state detector together with appropriated electronics \cite{Cho_pmb_2013}. Since this technique is quite expensive and requires specialized labor, it has been proposed  an alternative approach to reconstruct X-ray fluence spectra from its attenuation curves data.

In its pioneer paper, L. Silberstein proposed a method based on the Laplace Transform to reconstruct X-ray spectra using an analytical approximation to the attenuation curve  \cite{Silberstein1933}. In 1936, G. E. Bell discusses Silverstein technique and tests its performance using experimental curves \cite{Bell1936}.  Some decades later, J. W. Twidell presented a computer code to reconstruct iteratively X-ray spectra by error minimization between experimental and numerical attenuation data \cite{Twidell1970}. In 1983, Kramer \& Von Seggern published two reviews of the main techniques to solve the X-ray spectra reconstruction problem and showed results obtained by the least square method \cite{Kramer1983a,Kramer1983b}. One year later Rubio \& Mainardi applied the Laplace Transform to reconstruct X-ray spectra with characteristic lines. \cite{Rubio_pmb_1984}. 

In the last years, several studies have been published dealing with experimental and numerical aspects of the X-ray spectrum problem \cite{Delgado2009,Raspa2010,Ali_medphys_2012a,Ali_medphys_2012b}. Specifically, in 1998 Nisbet et al proposed the using of Classical Simulated Annealing (CSA) algorithm to reconstruct X-ray spectra with energy between $6\,\mbox{MeV}$ and $25\,\mbox{MeV}$ \cite{Nisbet1998} by the minimization of an objective function. Typically, this function is defined as the error between the experimental and modeled data added to a regularization function through a factor $\lambda$. However, it was not clearly shown how the regularization function was built neither how each new solution has been generated from the previous ones.

As it is widely known, simulated annealing is a stochastic optimization algorithm which has been employed to  solve non convex maximization/minimization problems. Potential solutions are iteratively generated through a suitable visitation distribution and whether accepted or not according to a predefined criterion. Its classical version (CSA) was formulated based on the Metropolis acceptance criterion \cite{Metropolis1953} combined to a geometric or logarithmic temperature cooling schedule \cite{Kirkpatrick1983} and adopting a Gaussian visitation distribution. It has been mathematically shown that the logarithmic cooling temperature ensures the global minimum convergence \cite{Geman1984}. Since such cooling schedule takes excessive computational time, a faster algorithm was proposed, the Fast Simulated Annealing (FSA), which accomplishes a visitation distribution based on Cauchy probability density function with the cooling schedule as inverse of the iteration \cite{Szu1987}. 

By the introducing of two new parameters, $q_a$ and $q_v$, Tsallis \& Stariolo designed the Generalized Simulated Annealing (GSA), which generalizes the Metropolis acceptance criterion, the visitation distribution and the cooling schedule, retrieving CSA and FSA as particular cases \cite{Tsallis1996}.  Practical problems such as protein structure configurations and the electrical impedance tomography inverse problem have also been tackled using the simulated annealing algorithms \cite{Hansmann2010,Menin2013}.

Here we introduce an algorithm based on GSA combined with a suitable first derivative regularization function to solve the X-ray spectrum reconstruction inverse problem. A \textit{C} language computer code has been implemented and numerical experiments have been carried out to survey the algorithm performance with respect to two main parameters, $\lambda$ and $\tilde{q}_v$, which we conveniently define as $\tilde{q}_v = q_v - 1$. Results show that our algorithm has significant performance improvement compared to the CSA version without regularization function and is able to retrieve the X-ray fluence distribution with high accuracy. Moreover, results allow us to suggest a range of $\lambda > 10^{-2}$ and $\tilde{q}_v > 0.6$  that improves the algorithm performance.

The paper is organized as follows: in Sec. \ref{theoretical_exp_methods}, we state the X-ray fluence distribution reconstruction inverse problem and present the generalized simulated annealing algorithm. Also, we discuss the computational implementation and present the numerical experiments. In Sec. \ref{results_discussion}, we show and discuss the results and, finally, in Sec. \ref{conclusion}, we conclude stressing the effectiveness of our method.

\section{Experimental}
\label{theoretical_exp_methods}

In this section, we present the X-ray spectrum reconstruction inverse problem formulation and our proposal for a regularization function. Then, we introduce the generalized simulated annealing algorithm fundamentals. Finally, we discuss the algorithm implementation and present the numerical experiments.

\subsection{X-ray spectrum reconstruction inverse problem}
\label{inverse_problem}

\begin{figure}[h]
\begin{center}
\subfigure[]{\includegraphics[scale=0.35]{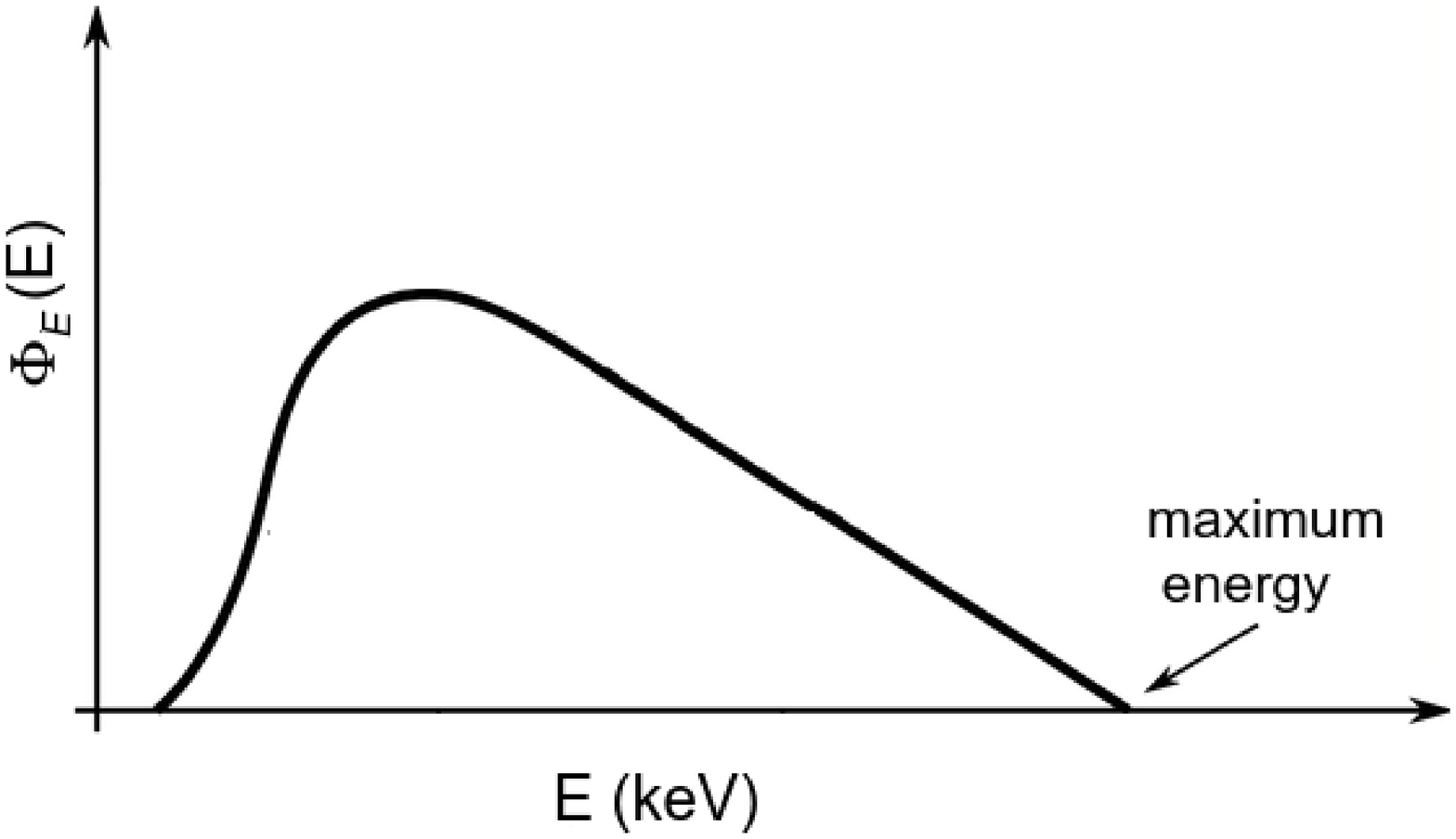}}
\subfigure[]{\includegraphics[scale=0.53]{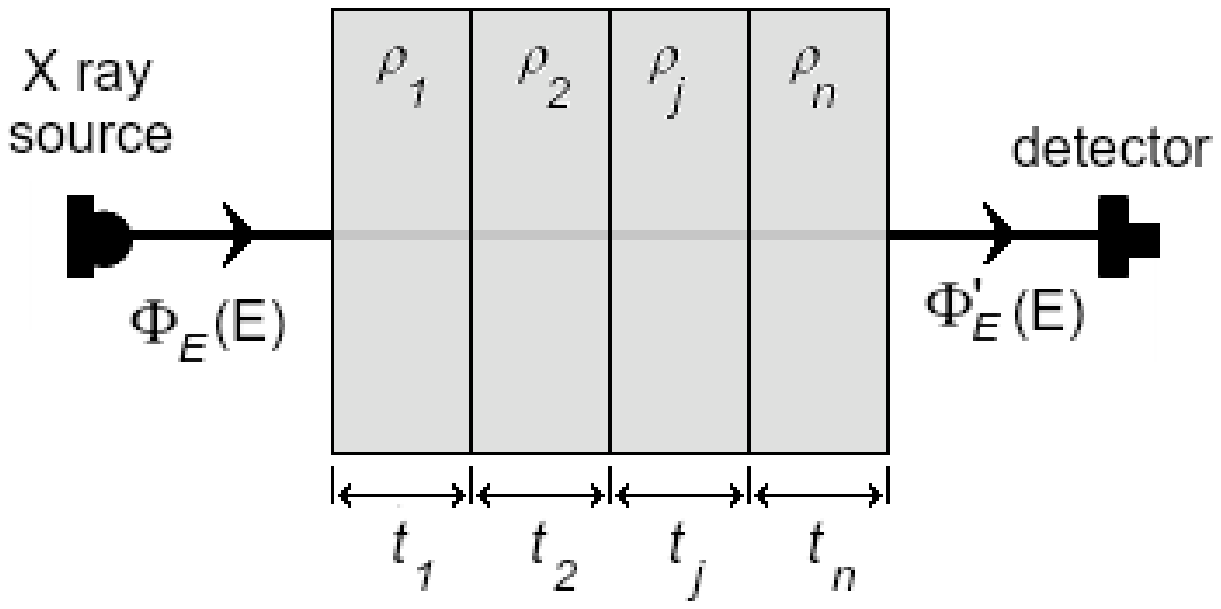}}
\end{center}
\caption{(a) Typical diagnostic X-ray spectrum and (b) a X-ray beam  with incident fluence distribution $\Phi_E(E)$ crossing $n$ attenuation plates with thickness $t_1, t_2, \cdots, t_n$ and densities $\rho_1, \rho_2, \cdots, \rho_n$. The attenuated fluence distribution is $\Phi'_E(E)$.}
\label{fig_spectrum_attenuation}
\end{figure}

Consider a X-ray beam with unattenuated fluence distribution $\Phi_E(E)$, as shown in Fig. (\ref{fig_spectrum_attenuation}-a). After crossing $n$ attenuation plates with thickness $t_1, t_2, \cdots, t_n$ and densities $\rho_1, \rho_2, \cdots, \rho_n$, as shown in Fig. (\ref{fig_spectrum_attenuation}-b), the attenuated fluence distribution $\Phi'_E(E)$ is

\begin{equation}
\Phi'_E(E) = \Phi_E(E) \exp\left[-\left(\mu/\rho \right)\sum\limits_{j=0}^{n}\rho_jt_j\right],
\label{atenuation}
\end{equation}

\noindent where $\left(\mu/\rho\right)$ is the mass attenuation coefficient, which depends on $E$. The relative transmission $T_n$ after the X-ray beam cross $n$ attenuation plates is

\begin{equation}
T_n = \frac{\int_0^{E_{\rm max}}\Phi_E(E)(\mu_{\rm en}/\rho)\exp\left[-(\mu/\rho)\sum_{j=0}^n\rho_jt_j\right]EdE}{\int_0^{E_{\rm max}}\Phi_E(E)(\mu_{\rm en}/\rho)EdE},
\label{trasmission}
\end{equation}

\noindent where $E_{\rm max}$ is the spectrum energy upper limit and $(\mu_{\rm en}/\rho)$ is the mass energy-absorption coefficient, which depends on $E$. 

Then, one defines the X-ray spectrum reconstruction direct problem as the determination of the transmission curve ${T_n}$ from a known fluence distribution ${\Phi_E}$ and the correspondent inverse problem as the determination of the fluence distribution ${\Phi_E}$ from its transmission curve $T_n$. Discretizing the continuum fluence distribution $\Phi_E(E)$ in a vector ${\bf\Phi} = \left[\phi^{(0)},\phi^{(1)}, \cdots,\phi^{(k-1)}\right]$ that represents a histogram with $k$ energy bins, the integrals of Eq. (\ref{trasmission}) can be replaced by summations and the transmission becomes the vector ${\bf T} = \left[T^{(0)},T^{(1)}, \cdots, T^{(m-1)}\right]$, with components

\begin{equation}
T^{(n)} = \frac{\sum\limits_{s=0}^{k-1}a_{n,s}\phi^{(s)}}{\sum\limits_{s=0}^{k-1}a_{0,s}\phi^{(s)}}, \quad\mbox{for}\quad n = 0,1,\cdots,m-1,
\label{transmission_discret}
\end{equation}

\noindent where $m$ is the number of attenuation measurements and

\begin{equation}
a_{n,s} = \int_{E(s)}^{E(s+1)} (\mu_{\rm en}/\rho)(E)\exp\left[-(\mu/\rho)(E)\sum\limits_{j=0}^{n}\rho_jt_j\right]EdE.
\label{integral}
\end{equation}

Thus, in our discretized X-ray spectrum reconstruction inverse problem, we know the transmission data ${\bf T}$ and we must determine the fluence distribution ${\bf\Phi}$. Comparing the experimental transmission data ${\bf S} = \left[S^{(0)}, S^{(1)}, \cdots, S^{(m-1)}\right]$ with the numerical one ${\bf T} = \left[T^{(0)},T^{(1)}, \cdots, T^{(m-1)}\right]$ generated from a trial fluence distribution ${\bf\Phi}_{\rm trial}$, we define an objective function as

\begin{equation}
f\left({\bf\Phi}_{\rm trial}\right) = \frac{1}{m}\sum\limits_{n=0}^{m-1}\left[T^{(n)} - S^{(n)}\right]^2 + \lambda R({\bf\Phi_{\rm trial}}),
\label{objective_function}
\end{equation}

\noindent where $R ({\bf\Phi_{\rm trial}})$ is a regularization function and $\lambda$ is the regularization parameter.

The suitable choice to the regularization function $R({\bf\Phi_{\rm trial}})$ is of fundamental importance. In our algorithm, it has been constructed to guarantee the smoothness of the spectrum penalizing large fluence differences between two neighboring components of ${\bf\Phi}_{\rm trial}$. Thus, we have considered it as the first derivative of the standardized trial fluence distribution

\begin{equation}
R({\bf\Phi}_{\rm trial})= \frac{1}{k-2}\sum\limits_{s=1}^{k-1}\left|\hat{\phi}_{\rm trial}^{(s)}-\hat{\phi}_{\rm trial}^{(s-1)}\right|,
\label{regularization_function}
\end{equation}

\noindent where $\hat{\phi}_{\rm trial}^{(s)}$ are the elements of the standardized trial fluence

\begin{equation}
{\hat{\bf\Phi}}_{\rm trial} = \frac{{\bf\Phi}_{\rm trial}}{\max[{\bf\Phi}_{\rm trial}]}.
\label{normalized_fuence}
\end{equation}

Our propose for such standardization is important to makes the regularization function $R ({\bf\Phi_{\rm trial}})$ to have the same order magnitude of the absolute mean square error $\sum\limits_{n=0}^{m-1}\left[T^{(n)} - S^{(n)}\right]^2/m$ of Eq. (\ref{objective_function}) and facilities the choice of the regularization parameter $\lambda$, which represents the threshold between the data fitting and the spectrum smoothness. Moreover, since $T^{(n)}$ and $S^{(n)}$ are relative transmission, our standardization of the trial fluence distribution makes the objective function Eq. (\ref{objective_function}) dimensionless.

Summarizing, in the X-ray reconstruction spectrum inverse problem, one must find out a fluence distribution that yields the objective function global minimum. Mathematically, the solution of the optimization problem  ${\bf\Phi}_{\rm trial}^*$ must satisfy 

\begin{equation}
f({\bf\Phi}_{\rm trial}^*) < f({\bf\Phi}_{\rm trial}), \quad\forall\quad {\bf\Phi}_{\rm trial} \neq {\bf\Phi}_{\rm trial}^*,
\label{sol_inverse_prob}
\end{equation}

\noindent which we are concerned in the following.

\subsection{Generalized Simulated Annealing Algorithm}
\label{gsa}

The algorithm starts from a random initial solution. At each iteration, a new guess solution ${\bf\Phi}_{\rm new}$ is generated and compared with the previous one ${\bf\Phi}_{\rm old}$. Before describing the way new solutions are generated, let us address the acceptance criterion. According to Metropolis criterion, the probability $p({\bf\Phi}_{\rm old} \leftarrow {\bf\Phi}_{\rm new})$ to accept the new solution is 

\begin{equation}
\begin{array}{lcl}
p({\bf\Phi}_{\rm old} \leftarrow {\bf\Phi}_{\rm new}) &=& \left\{ \begin{array}{ll} 1, & \mbox{if}\quad \Delta f \leq 0\\
e^{-\Delta f/T_a}, &\mbox{if}\quad \Delta f > 0\end{array}\right.\end{array}
\label{metropolis}
\end{equation}

\noindent where $\Delta f = f({\bf\Phi}_{\rm new}) - f({\bf\Phi}_{\rm old})$ is the variation between the new and old solutions objective function and $T_a$ is the acceptance temperature, which corresponds to a stochasticity control parameter. 

To generate a new solution ${\bf\Phi}_{\rm new}$, each component $\phi_{\rm old}^{(s)}$ of the current solution ${\bf\Phi}_{\rm old}$ is modified, one at a time, generating a new component $\phi_{\rm new}^{(s)}$ through the visitation distribution

\begin{equation}
\phi_{\rm new}^{(s)} = \phi_{\rm old}^{(s)} + T_v G_{\tilde{q}_v}(\phi), \quad\mbox{for}\quad s = 0,1,\cdots,k-1,
\label{visitation}
\end{equation}

\noindent where $T_v$ is the visitation temperature, which is related to the visitation length jump,  and

\begin{equation}
G_{\tilde{q}_v}(\phi) = \frac{\Gamma\left(1/\tilde{q}_v\right)\sqrt{\tilde{q}_v}}{\Gamma\left(1/\tilde{q}_v-1/2\right)\sqrt{2\pi}} \exp_{-\tilde{q}_v}\left[-\frac{1}{2}\phi^2\right]
\label{qgaussian}
\end{equation}

\noindent is the standard $\tilde{q}$-Gaussian probability distribution function (pdf) within

\begin{equation}
\begin{array}{lcl}
\exp_{\tilde{q}}(x) &=& \left\{\begin{array}{ll}
0, & \mbox{se}\quad \tilde{q}x < -1 \\
\lim_{\tilde{q}' \rightarrow \tilde{q}} \left(1+\tilde{q}'x\right)^{1/\tilde{q}'},  & \mbox{se}\quad \tilde{q}x \geq -1
\end{array}\right.
\end{array}
\label{q_exp}
\end{equation}

\noindent being the generalized exponential function \cite{Martinez2009}.

The attractive feature of the generalized Gaussian visitation is its versatility. The suitable choice of $\tilde{q}_v$ parameter allows the algorithm transit from a global search to a local one and vice-versa. The 2nd moment of the $\tilde{q}$-Gaussian pdf (\ref{qgaussian}) is only finite for $\tilde{q}_v < 2/3$, which allows us to infer that for $\tilde{q}_v \approx 0.7$, one has the best agreement between local and global visitation \cite{Martinez2009}. Tsallis \& Stariolo, for example, estimated that the GSA best performance is reached to $\tilde{q}_v \approx 1.7$ \cite{Tsallis1996},  while Deng et al proposed an approach to generate $\tilde{q}$-Gaussian deviate and show numerically  that the optimal GSA performance is to $\tilde{q}_v \approx 1.3$ \cite{Deng2005}.

The stochastic  search is controlled by the temperature cooling schedule. The process starts with high temperatures allowing a wide spread visitation and high probability to accept a new solution that increases the objective function. As the process approaches the end, the visitation becomes more local and the probability to accept bad new solutions tends to vanish. We have chosen two schedules to decrease the temperatures $T_a$ and $T_v$ as a function of the iteration $t$,

\begin{equation}
T_a^{(t)}= \alpha^t T_a^{(0)},
\label{Tacooling}
\end{equation}
\noindent where $\alpha \in (0,1)$ is a cooling rate, and

\begin{equation}
T_v^{(t)} = T_v^{(0)}\frac{\ln_{\tilde{q}_v}(2)}{\ln_{\tilde{q}_v}(t+1)},
\label{Tvcooling}
\end{equation}

\noindent with $\ln_{\tilde{q}}(x) = \lim_{\tilde{q}' \rightarrow \tilde{q}} \left[(x^{\tilde{q}'}+1)/\tilde{q}'\right]$ being the generalized logarithm function, which is the inverse of the presented generalized exponential function Eq. (\ref{q_exp}).

It is important to note that as $\tilde{q}_v \rightarrow 0$, Eq. (\ref{qgaussian})  and Eq. (\ref{Tvcooling}) become, respectively, the usuals Gaussian distribution and logarithm cooling schedule, retrieving the  CSA. Also it is possible verify that for $\tilde{q}_v \rightarrow 1$ we obtain the FSA with Cauchy visitation distribution and inverse of iteration cooling schedule.

One of the present paper contributions is the initial temperatures setting approach. To guarantee initial large probability to detrap from a local minimum basin the initial acceptance temperature $T_a^{(0)}$ must be greater than the depth of deepest local minimum landscape. To estimate $T_a^{(0)}$,  an initial random search is performed in which $\mu$ solutions ${\bf\Phi}_1, {\bf\Phi}_2, \cdots, {\bf\Phi}_{\mu}$ are generated and its correspondent objective functions $f({\bf\Phi}_1), f({\bf\Phi}_2), \cdots, f({\bf\Phi}_{\mu})$ are calculated. Evaluating the greatest difference of the objective function $\Delta f_\mu = \max\left[f({\bf\Phi})\right] - \min\left[f({\bf\Phi})\right]$, we define

\begin{equation}
T_a^{(0)} = -\frac{\Delta f_\mu}{\ln(p_0)},
\label{Ta_0}
\end{equation}

\noindent where $p_0 \in (0,1)$ is a predefined initial probability to accept new solutions that increase the objective function and is set near to $1$ to give the algorithm initial high probability to escape from local minima basin.  

To ensure an efficient wide landscape exploration, the initial visitation temperature, $T_v^{(0)}$, must be greater than the largest actual spectrum fluency. To simplify the $T_v^{(0)}$ setting, we modeled the actual spectrum shape as a triangle with base $E_{\rm max} - E_{\rm min}$ and high $T_v^{(0)}$, where $E_{\rm min}$ and $E_{\rm max}$ are the minimum and maximum spectrum energy. As the collision air kerma $K_{\rm air}$ can be experimentally measured and it represents the area underneath of the modeled spectrum (triangle), we set

\begin{equation}
T_v^{(0)} = \frac{2K_{\rm air}}{E_{\rm max}-E_{\rm min}}.
\label{Tv_0}
\end{equation}

Summarizing, consider the following GSA algorithm, being $t_{\rm max}$ the maximum number of iterations and $r$ the number of new trial solutions at each iteration.

\begin{enumerate}
		\item set the simulation parameters: $t_{\rm max}$, $r$, $\alpha$, $q_v$ e $\lambda$;
		\item set the initial temperatures $T_a^{(0)}$ and $T_v^{(0)}$ using Eqs. (\ref{Ta_0}) and (\ref{Tv_0});
		\item generate a random initial solution ${\bf\Phi}_{\rm old}$;
		\item evaluate the objective function $f\left({\bf\Phi}_{\rm old}\right)$ using Eq. (\ref{objective_function});
		\item for $t$ from $1$ to $t_{\rm max}$:
		\begin{enumerate}
		\item repeat $r$ times the following steps:
			\begin{enumerate}
			\item generate a new solution ${\bf\Phi}_{\rm new}$ using Eq. (\ref{visitation});
			\item evaluate the new objective function $f\left({\bf\Phi}_{\rm new}\right)$ using Eq. (\ref{objective_function});
			\item Considere the Metropolis criterion:
			
			$\bullet$ if $f\left({\bf\Phi}_{\rm new}\right) \leq f\left({\bf\Phi}_{\rm old}\right), \quad \mbox{then}\quad {\bf\Phi}_{\rm old} = {\bf\Phi}_{\rm new}$;
			
			$\bullet$ else, if $e^{\left[f\left({\bf\Phi}_{\rm old}\right) - f\left({\bf\Phi}_{\rm new}\right)\right]/T_a} > X \sim U(0,1), \quad \mbox{then}\quad {\bf\Phi}_{\rm old} = {\bf\Phi}_{\rm new}$;
			\end{enumerate}
		\item cooling down the temperatures $T_a$ and $T_v$ using Eqs. (\ref{Tacooling}) and (\ref{Tvcooling});
		\end{enumerate}
	\item output results
\end{enumerate}

\subsection{Implementation and numerical experiments}
\label{implementation_tests}

A \textit{C} language computer code to solve the X-ray spectrum reconstruction inverse problem have been implemented and numerical tests were carried out to assess its performance. Since our contributions are mainly related with a suitable regularization function and the use of GSA instead of CSA, the numerical experiments have dealt specifically to the algorithm accuracy with respect to the parameters $\tilde{q}_v$ and $\lambda$. 

The implemented computer code receives, as input data, the collision air kerma, $K_{\rm air}$, the minimum and maximum spectrum energies, $E_{\rm min}$ and $E_{\rm max}$, and the experimental transmission data, ${\bf S}$. Thus, the GSA algorithm (Sec. \ref{gsa}) is throughout performed $N$ times starting from different random initial guess solutions, yields $N$ numerical spectra ${\bf\Phi}_1, {\bf\Phi}_2, \cdots,{\bf\Phi}_N$. The final solution is the numerical fluence distribution ${\bf\Phi_{\rm num}}$ defined as

\begin{equation}
\phi_{\rm num}^{(s)} = \frac{1}{N}\sum\limits_{i=1}^{N}\phi_i^{(s)} \qquad\mbox{for}\quad s = 0,1,\cdots,k-1,
\label{finalsolution}
\end{equation}

\noindent where $\phi_{\rm num}^{(s)}$ is the numerical mean fluence for each energy bin $s$ of ${\bf\Phi_{\rm num}}$ and $k$ is the number of energy bins. To generate the $\tilde{q}_v$-Gaussian deviate, we have used the generalized Box-M\"uller method \cite{Thistleton2007}.

\begin{figure}[h]
\centering\subfigure{\includegraphics[scale=0.52]{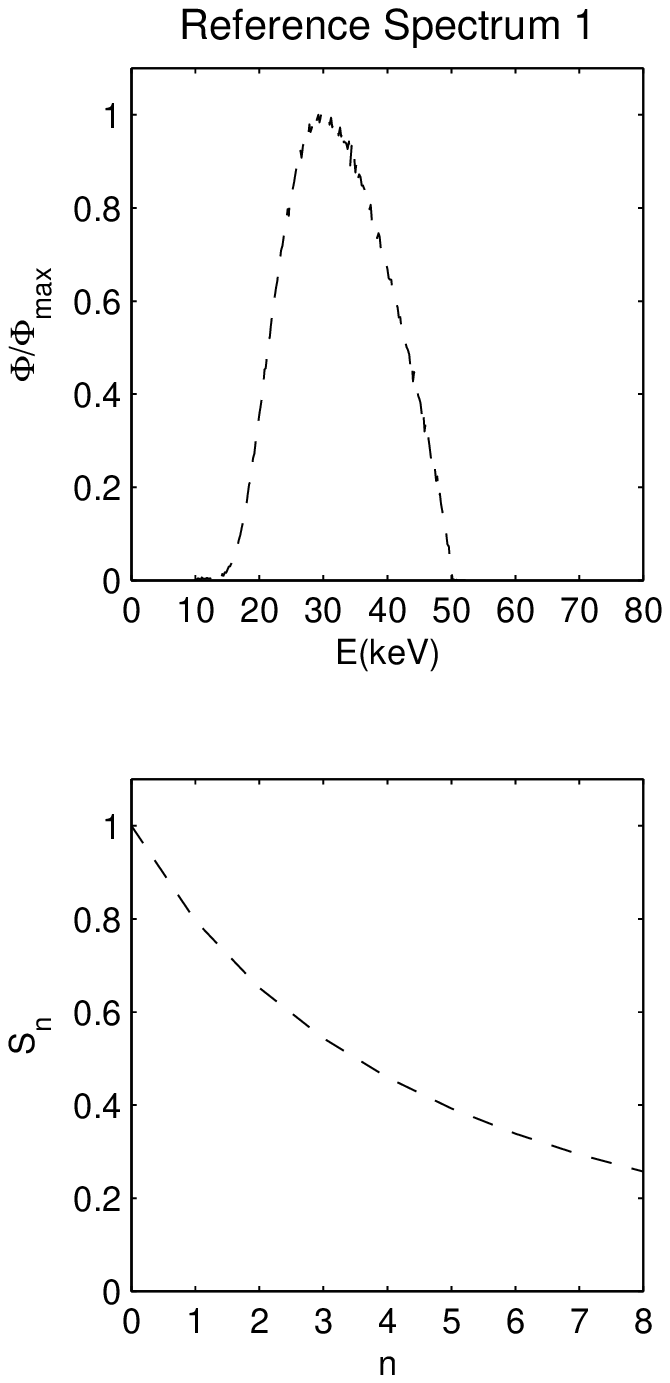}}
\centering\subfigure{\includegraphics[scale=0.52]{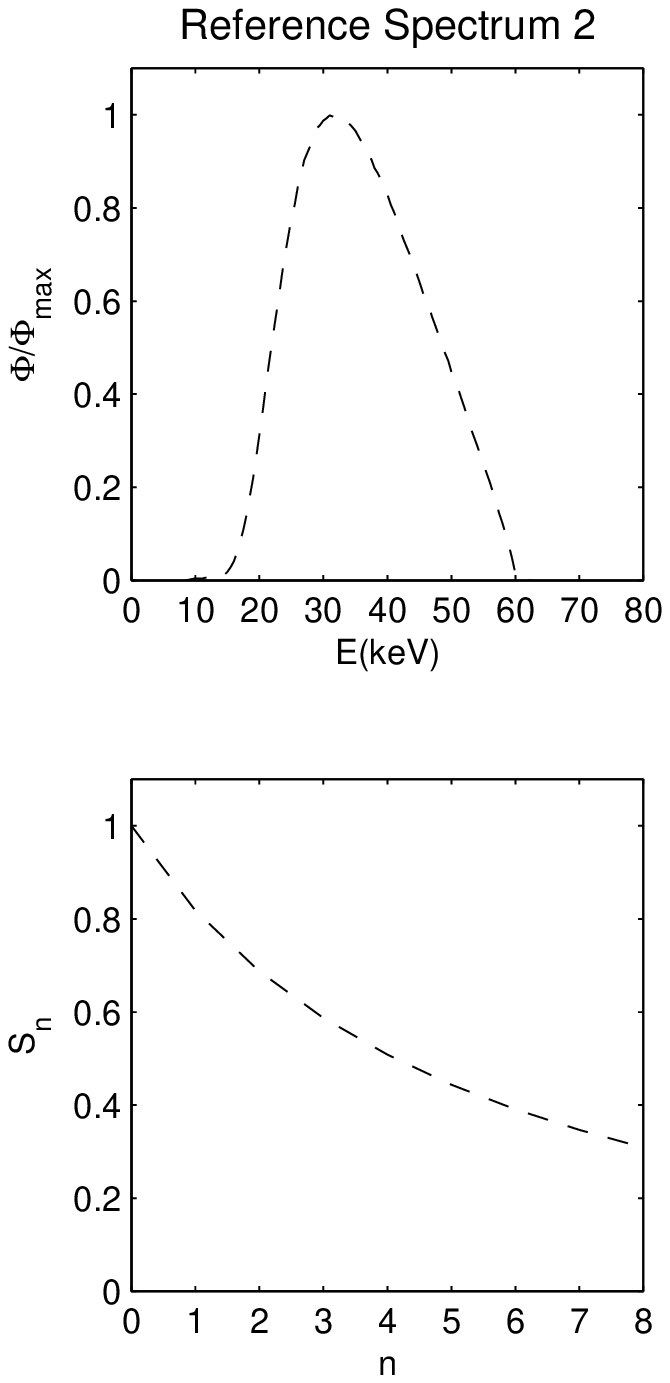}}
\centering\subfigure{\includegraphics[scale=0.52]{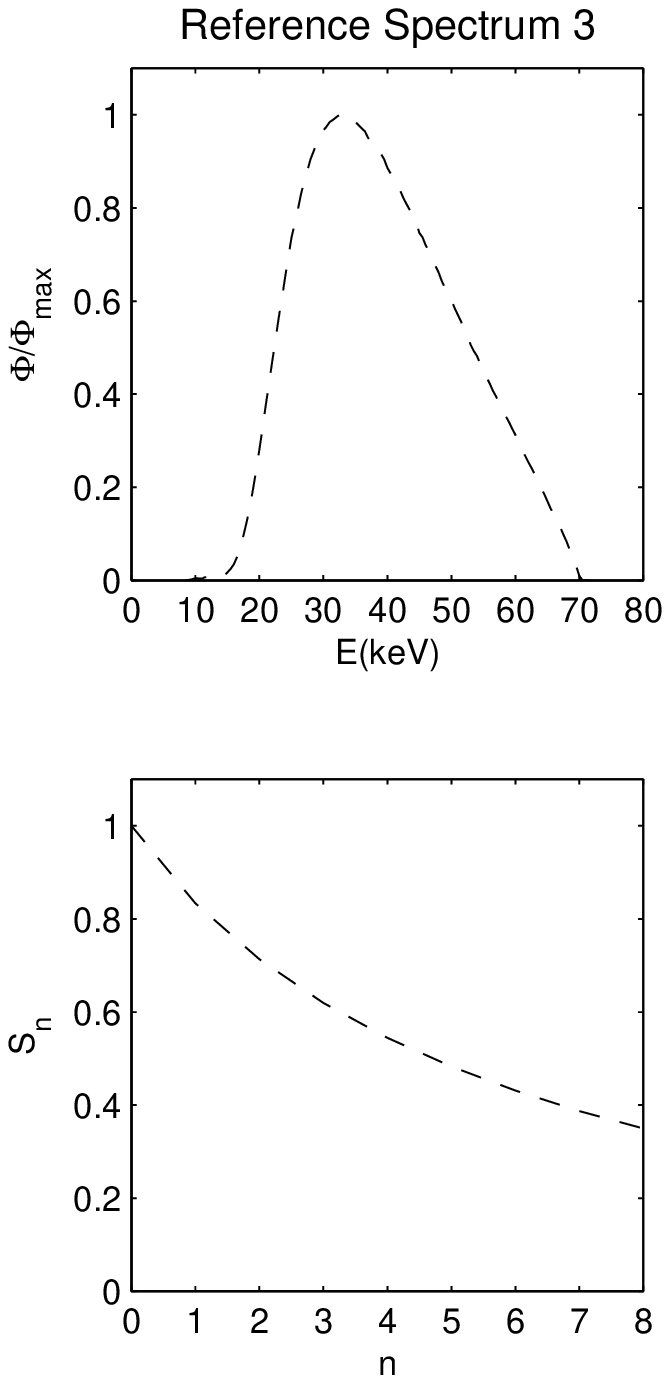}}
\caption{Normalized fluence distributions $\Phi/\Phi_{\rm max}$ and the correspondent transmission $S_n$ curves for the reference spectra.}
\label{actualspectra}
\end{figure}

Three experimental spectra data with peak voltages of $50\,\mbox{kVp}$, $60\,\mbox{kVp}$ and $70\,\mbox{kVp}$ were selected as reference to test the suitability of our algorithm \cite{Ankerhold2000}. The choice was made to avoid characteristic X-ray in addition to the bremsstrahlung spectrum. The reference spectra minimum energy, $E_{min}$, maximum energy, $E_{max}$, and number of energy bins $k$ are shown in Tab. \ref{tab_actual_spectra}. Attenuation curves for aluminum considering $m = 9$ different thickness were numerically generated and used as input data in our numerical experiments. The normalized fluence distributions for each reference spectra and its relative transmission curves are shown in Fig. (\ref{actualspectra}). 

To compare the numerical solution ${\bf\Phi}_{\rm num}$ to the actual one ${\bf\Phi}_{\rm actual}$, we have chosen the absolute mean error

\begin{equation}
\varepsilon = \frac{1}{k-1}\sum\limits_{s=0}^{k-1} \left|\hat{\phi}_{\rm actual}^{(s)} - \hat{\phi}_{\rm num}^{(s)}\right|.
\label{abserror}
\end{equation}

\noindent where $\hat{\phi}_{\rm actual}^{(s)}$ and $\hat{\phi}_{\rm num}^{(s)}$ are the component $s$ of the actual and numerical normalized fluence distributions given, respectively, by 

$${\hat{\bf\Phi}}_{\rm actual} = \frac{{\bf\Phi}_{\rm actual}}{\max[{\bf\Phi}_{\rm actual}]} \qquad\mbox{and}\qquad {\hat{\bf\Phi}}_{\rm num} = \frac{{\bf\Phi}_{\rm num}}{\max[{\bf\Phi}_{\rm num}]}$$

\begin{table}[h!]
\centering
\caption{X-ray reference spectra with its energy range and number of energy bins \cite{Ankerhold2000}}
\label{tab_actual_spectra}
\begin{tabular}{c|ccc}
Spectrum & $E_{min}\, (\mbox{kev})$ & $E_{max}\, (\mbox{kev})$ & $k$ \\
\hline
1 & 10 & 52 & 211 \\
2 & 9 & 62 & 107 \\
3 & 9 & 72 & 127

\end{tabular}
\end{table}

For each reference spectra, the program has run setting $N = 25, t_{\rm max} = 200$, $r = 50$ and $\alpha = p_0 = 0.9$ constants, but varying $\tilde{q}_v$ and $\lambda$ in the range $\tilde{q}_v = 0, 0.2, 0.4, \cdots, 2.0$ and $\lambda = 10^{-6}, 10^{-5}, \cdots, 10^4$. To each pair ($\tilde{q}_v$,$\lambda$), the numerical fluence distribution ${\bf\Phi_{\rm num}}$ was compared to the actual one ${\bf\Phi_{\rm actual}}$ yielding $11\times 11 = 121$ absolute mean errors calculated by Eq. (\ref{abserror}).

\section{Results and discussion}
\label{results_discussion}

\begin{figure}[h!]
\centering\subfigure{\includegraphics[scale=0.44]{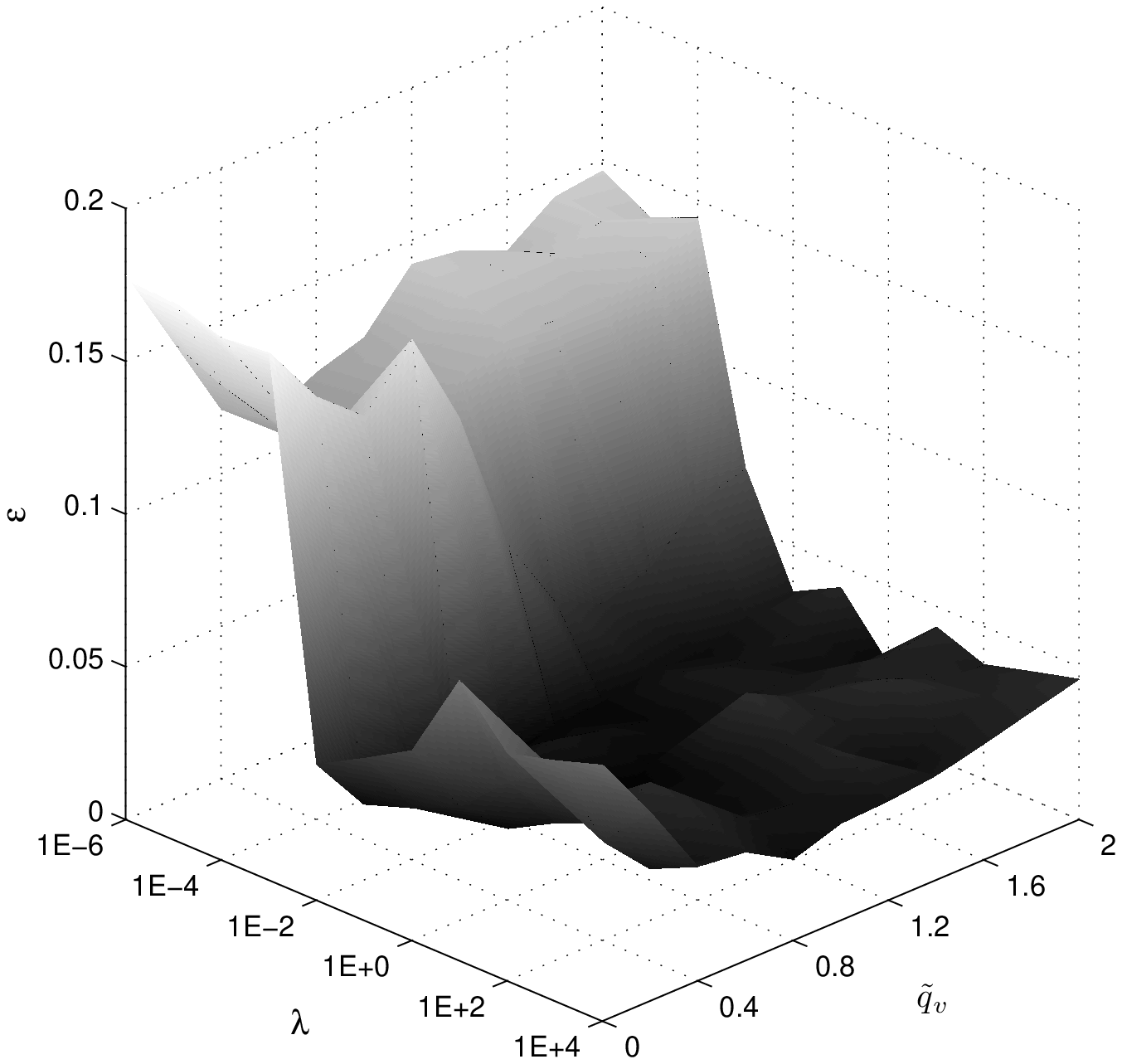}}
\hspace{1cm}
\centering\subfigure{\includegraphics[scale=0.38]{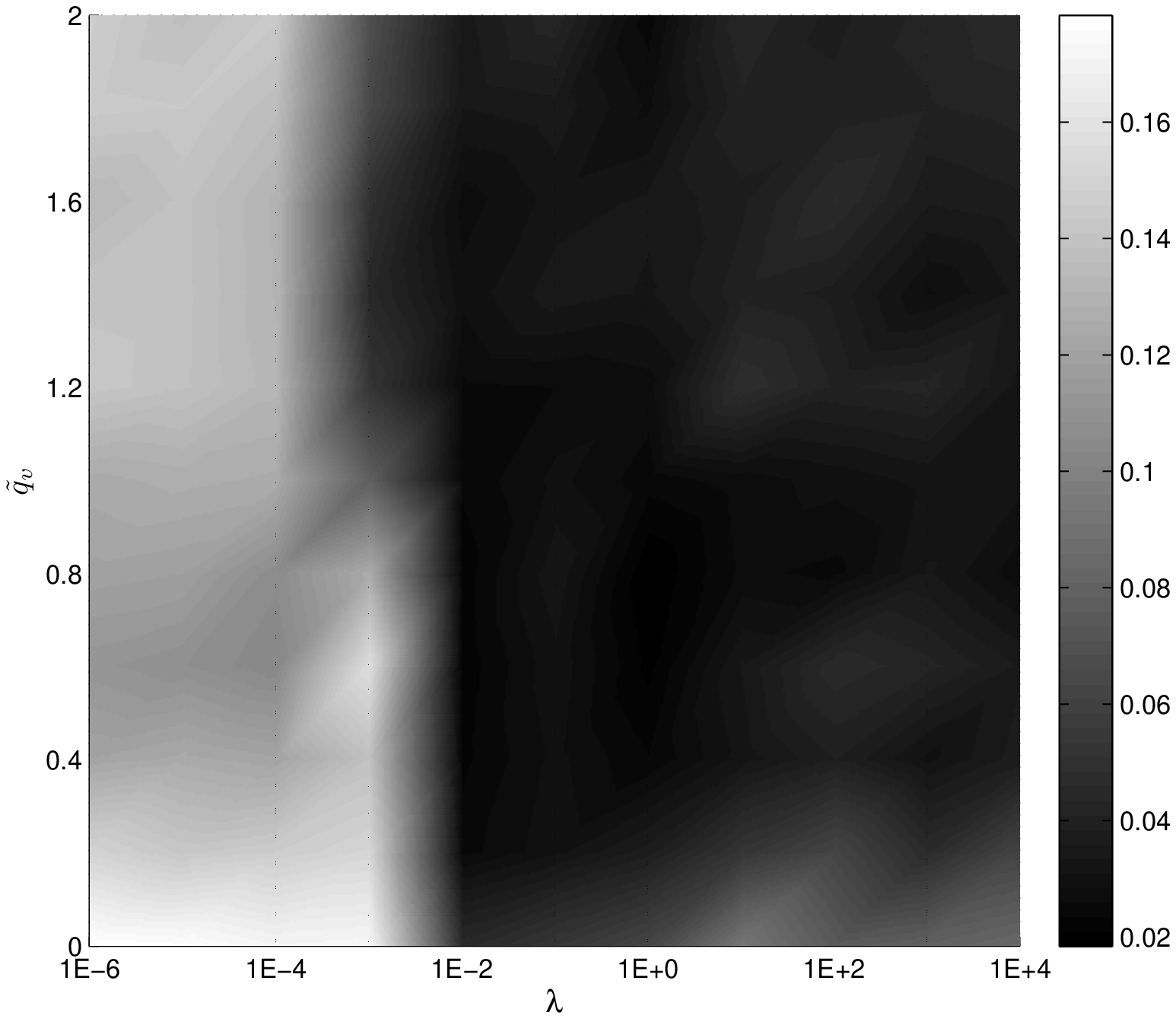}}\\ 
(a)\\
\bigskip
\centering\subfigure{\includegraphics[scale=0.44]{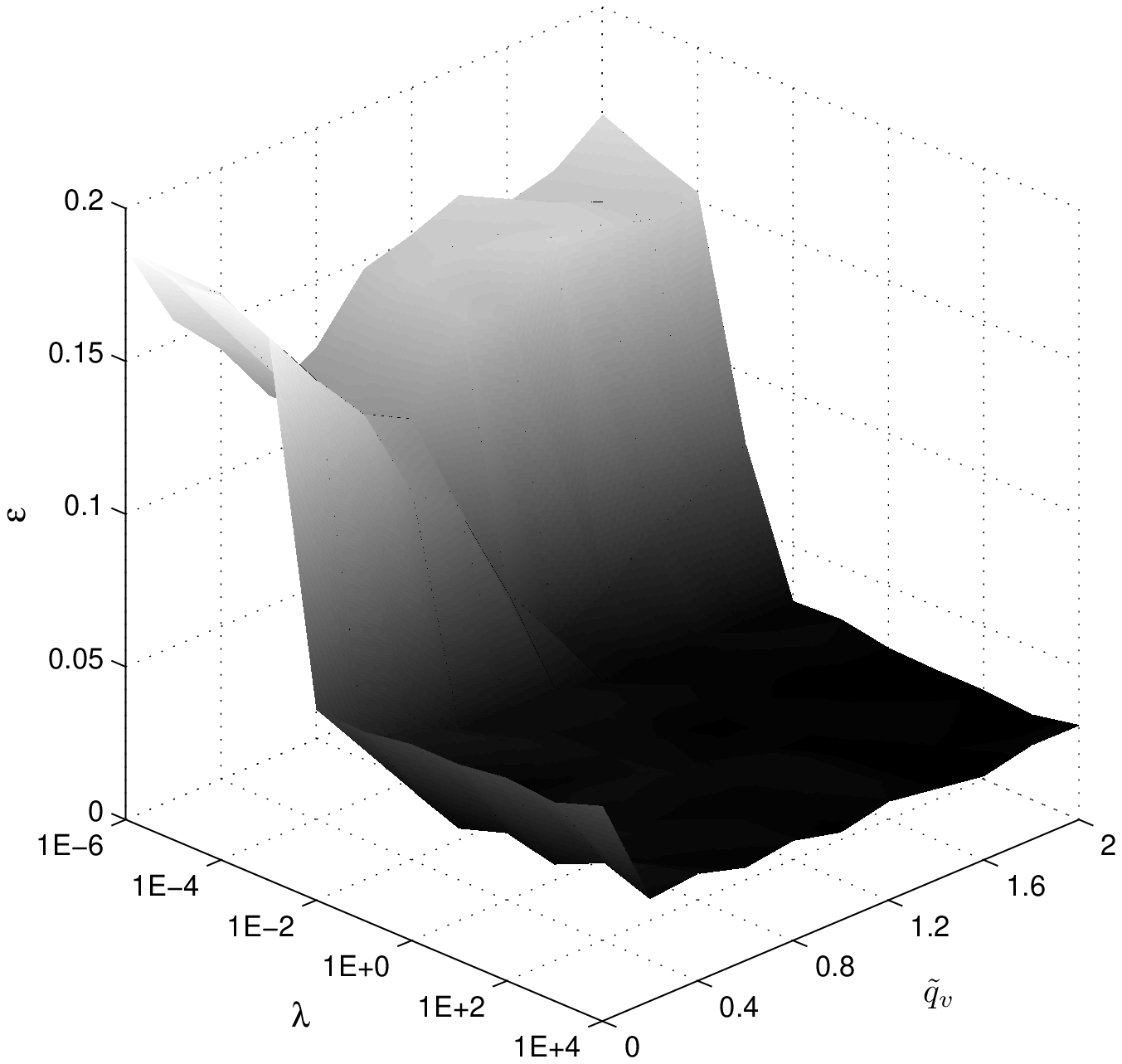}}
\hspace{1cm}
\centering\subfigure{\includegraphics[scale=0.38]{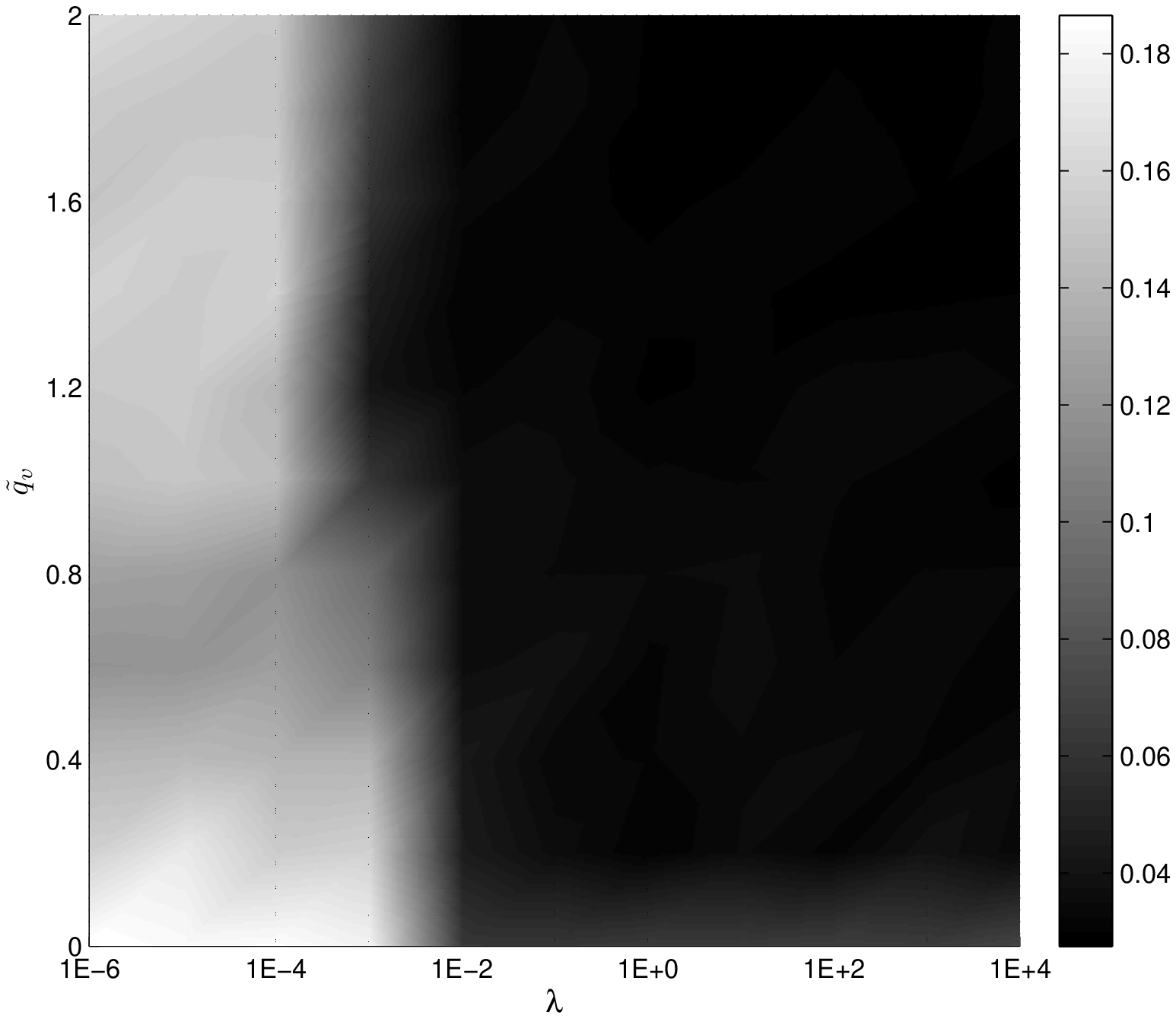}}\\ 
(b)\\
\bigskip
\centering\subfigure{\includegraphics[scale=0.44]{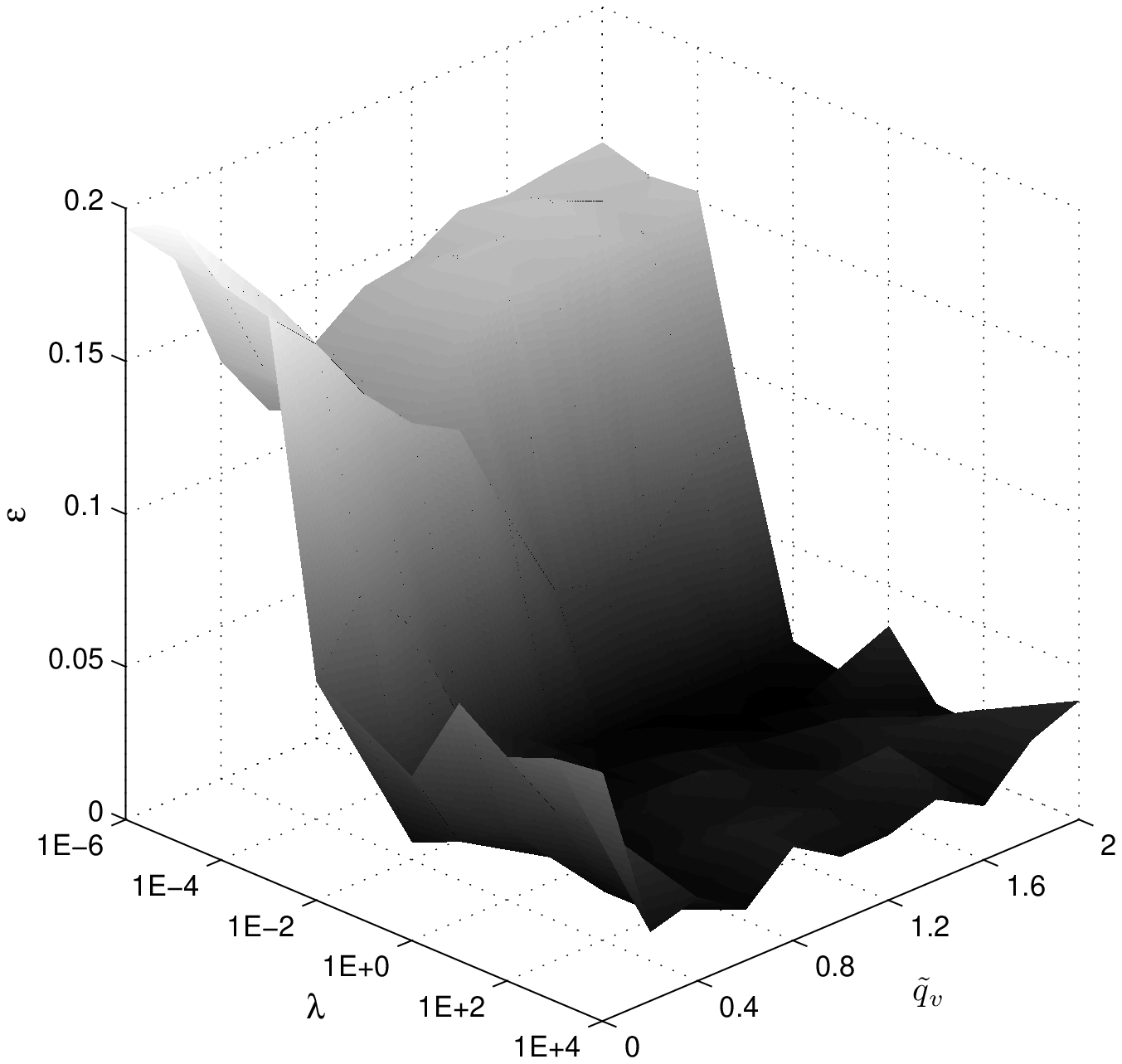}}
\hspace{1cm}
\centering\subfigure{\includegraphics[scale=0.38]{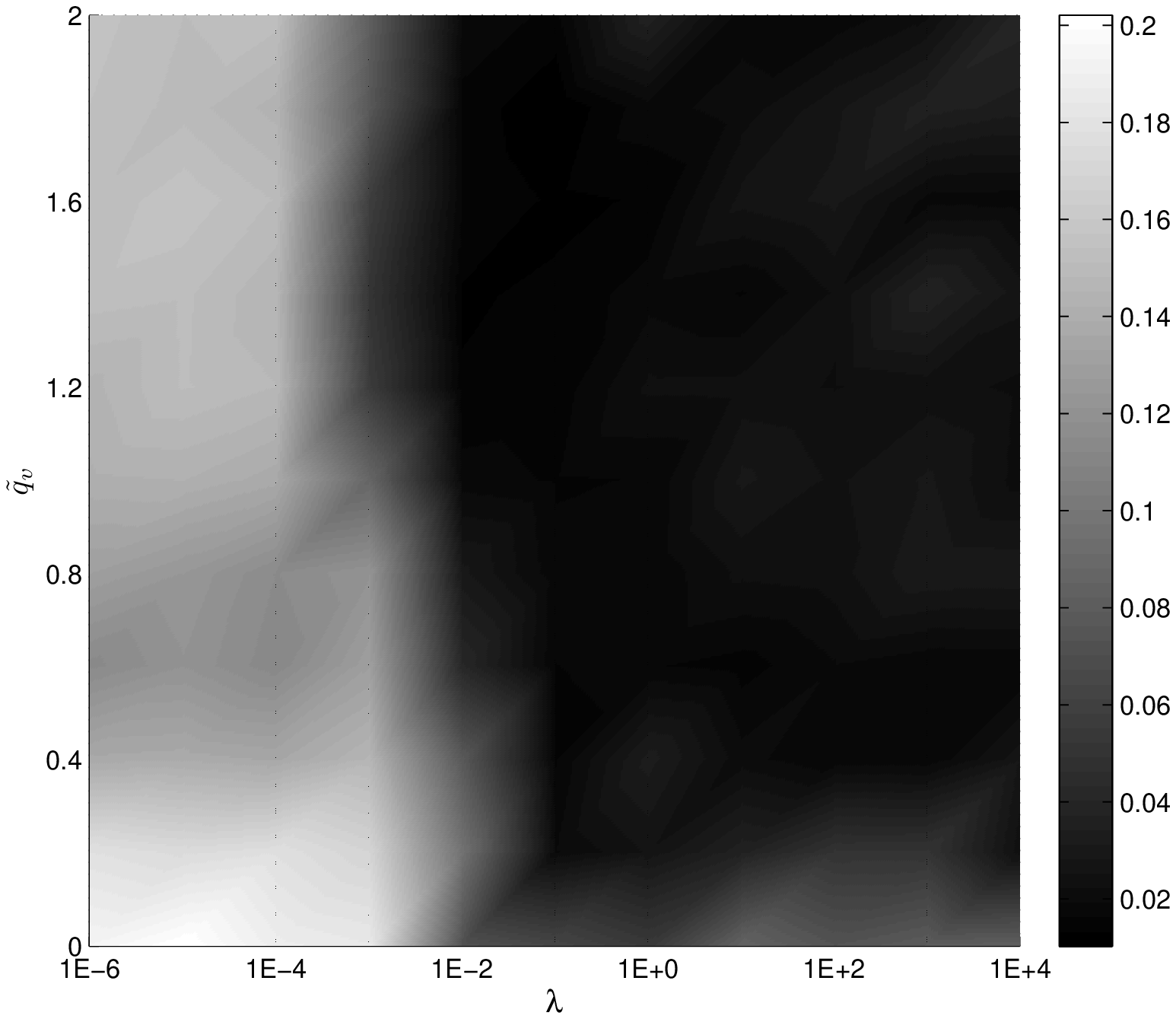}}\\ 
(c)
\caption{3-D surface viewing (left) and projection on $\lambda-\tilde{q}_v$ plane (right) of the absolute mean error $\varepsilon$ as function of $\tilde{q}_v$ and $\lambda$ for reference (a) Spectrum 1, (b) Spectrum 2 and (c) Spectrum 3. The error decreases drastically for $\lambda > 10^{-2}$ and for $\tilde{q}> 0.6$.}
\label{fig_surf}
\end{figure}

Now we present and analyze the main results obtained from the numerical experiments. First, we present in Fig. (\ref{fig_surf}) the absolute mean error $\varepsilon$ behavior as function of parameters $\lambda$ and $\tilde{q}_v$. Results does not allow us find out an unique ``optimal'' pair ($\lambda,\tilde{q}_v$). However it is notable that the error decreases strongly for $\lambda > 10^{-2}$ to all reference spectra, which corroborates the fundamental importance of our regularization function. Also, one can see considerable improvement on the results for $\tilde{q}_v > 0.6$. Indeed, as we have discussed in Sec. \ref{gsa}, the 2nd moment of pdf (\ref{qgaussian}) diverges for $\tilde{q}_v > 2/3$ and the visitation becomes more global and efficient. 

\begin{figure}[h!]
\begin{center}
\subfigure[]{\includegraphics[scale=0.70]{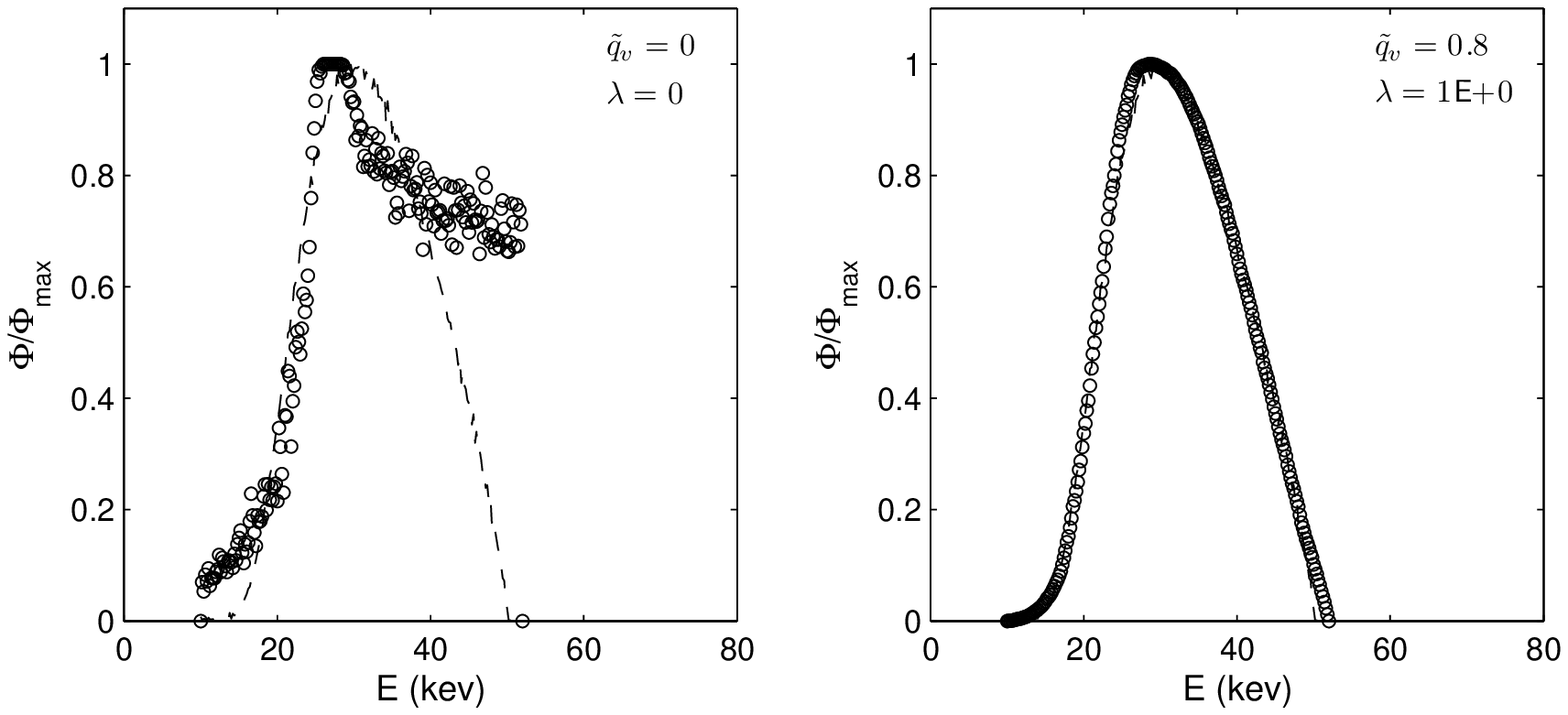}}
\subfigure[]{\includegraphics[scale=0.70]{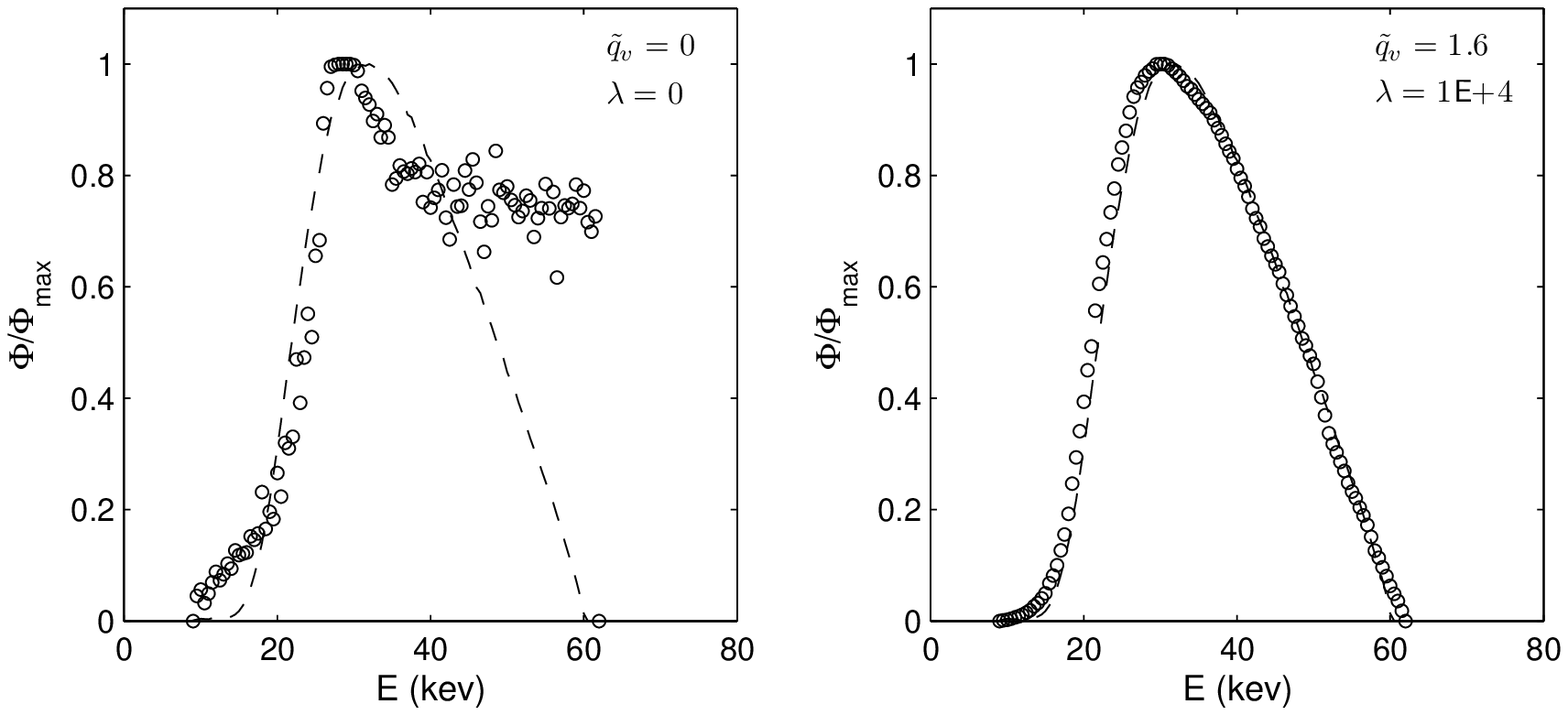}}
\subfigure[]{\includegraphics[scale=0.70]{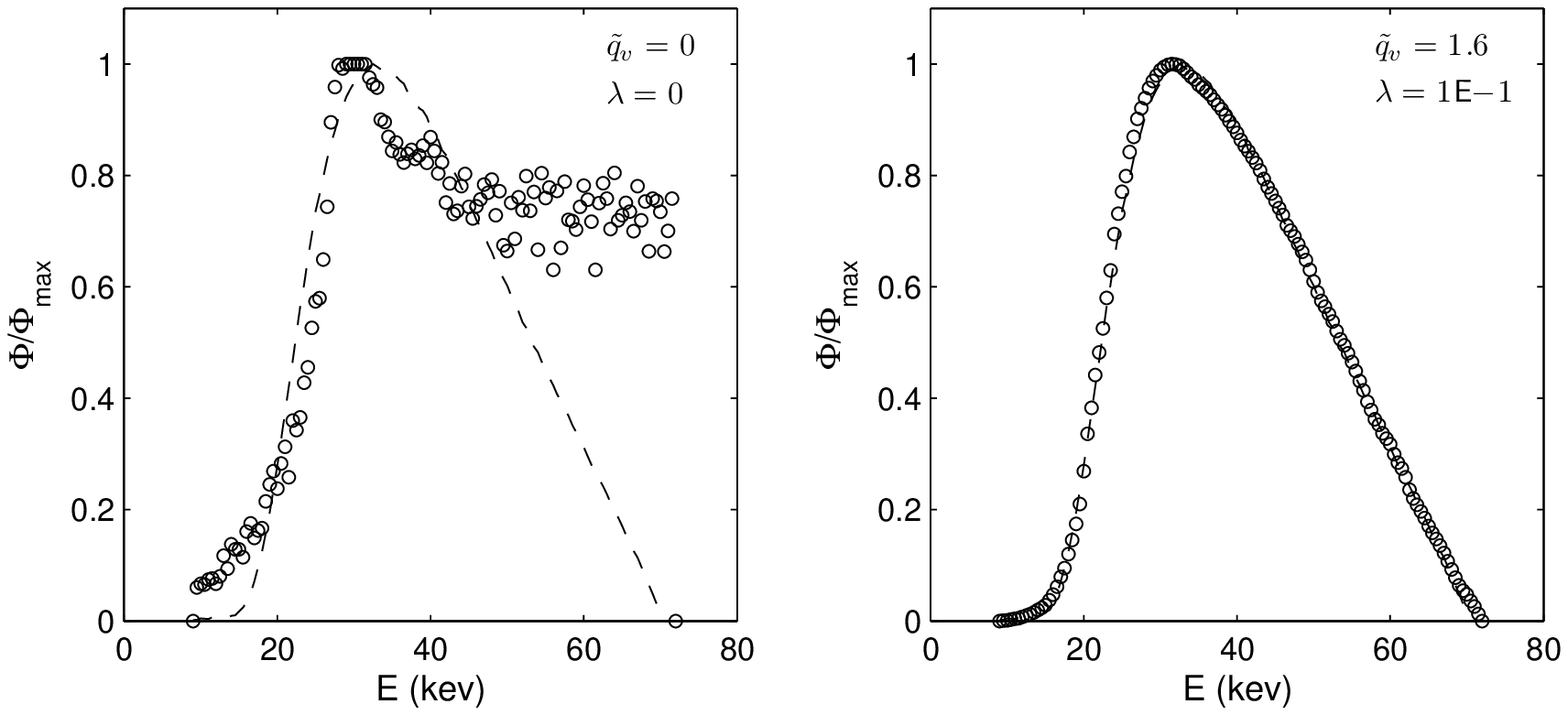}}
\caption{The actual fluence distribution (- - -) and the worst (left) and the best (right) numerical fluence distribution ($\circ$) for the reference (a) Spectrum 1, (b) Spectrum 2 and (c) Spectrum 3. It is evident the better fitness of GSA combined to our regularization function ($\lambda > 10^{-2}$ and $\tilde{q}> 0.8$) compared to CSA without regularization ($\lambda = 0$ and $\tilde{q} = 0$).}
\label{fig_spectra}
\end{center}
\end{figure}

Beyond that, to show how the numerical fluence distributions fit to the reference actual spectra,  we present in Fig. (\ref{fig_spectra}) the results obtained using CSA ($\tilde{q}_v =0$) without regularization function ($\lambda = 0$) and the best fluence distributions reached by GSA combined with our regularization function. Despite these best results have occurred for different values of $\tilde{q}_v$ and $\lambda$ for each reference spectra, it is quite evident that for $\tilde{q}_v > 0.6$ and $\lambda > 10^{-2}$ the algorithm provides excellent fitness to the actual data. 

Finally, although the Metropolis criterion might be also generalized, we have chosen its classical version since such generalization is still under investigation. Indeed, one year after the original paper of Tsallis \& Stariolo \cite{Tsallis1996},  Xiang et al proposed an rule to change the parameter $q_a$ to speed up the convergence of GSA  \cite{Xiang1997}. In 2009 Crokidakis et al have used a modified Metropolis criterion based on generalized Boltzmann-Gibbs entropy to analysis phase transitions of the Ising model \cite{Crokidakis2009} and three years later, Silva  et al \cite{Silva2012} have retrieved the detailed energy balance and locality of the generalized Metropolis criterion using the generalized operators \cite{Arruda2008}.

\section{Conclusions}
\label{conclusion}

A technique based on generalized simulated annealing algorithm combined to a first derivative regularization function to solve the X-ray spectrum reconstruction inverse problem was presented. A computer code that receives a X-ray attenuation data and returns an approximation for its fluence distribution was implemented. Numerical experiments using three reference spectra with $50\,\mbox{kVp}$, $60\,\mbox{kVp}$ and $70\,\mbox{kVp}$ were carried out to assess the algorithm performance mainly with respect to the parameters $\lambda$ and $\tilde{q}_v$. Results show high accuracy improvement for $\lambda > 10^{-2}$ and $\tilde{q}_v > 0.6$, corroborating the GSA better performance compared to CSA and showing the  fundamental importance of our smoothness regularization function. Moreover, our technique seems to be quite reliable since its numerical reconstruction fluence distributions present excellent fitness to the reference spectra. Finally, the presented technique is
promising to assist the tricky practical problem of determination of a X-ray spectrum beam from its attenuation data, which is considerably less expensive and easier to accomplish than the current direct measurement approaches.

\section*{Acknowledgements}

OHM thanks the IFSP and CAPES. ASM thanks the CNPq (305738/2010-0 and 485155-2013-3). AMC thanks the CNPq and FAPESP.

\end{document}